# Quantum register based on double quantum dots in semiconductor nanowires

*V. Vyurkov*[1,#], *L. Fedichkin*[2], *I. Semenikhin*[1], *D. Drozhzhin*[2], *K. Rudenko*[1], *V. Lukichev*[1]

[1)] Valiev Dept. of Physics and Technology, NRC "Kurchatov Institute," Moscow, Russia

[2)] Moscow Institute of Physics and Technology (National Research University), Dolgoprudny, Russia

[#]vyurkov.vv@mipt.ru

## ABSTRACT

An implementation of a universal solid-state quantum register based on electron space states in field-defined double quantum dots (a DQD possesses one electron in two adjacent tunnel bound dots) in an ultrathin semiconductor wire is discussed. To some extent, the structure resembles that of a field-effect transistor with multiple controlling electrodes (gates). Scalability is audible and it opens up a possibility of large-scale universal quantum computer fabricated by advanced silicon technology. Moreover, the structure could be developed into an ensemble quantum register where instead of single nanowire an array of them with common controlling electrodes and contacts is fabricated. That register is much more resistant against environment noise caused by phonons and random charges.

It is crucial that an individual qubit consists of *two* DQDs. The quantum information is encoded and processed inside the Hilbert subspace without charge transfer between dots. The filling factor of each quantum dot is permanently equal to 1/2. This guarantees a linear dynamics of qubits necessary for now existing quantum algorithms. Worth noting, the dynamics of qubits with altering charge state is more or less nonlinear due to interaction with surrounding dielectrics and metals (polaron effect). Even for weak interaction the accuracy of quantum operations becomes limited.

The basic two-qubit operations in the system are SWAP and √SWAP. The latter operation is universal as well as CNOT. The two-qubit operations are performed by Coulomb interaction. Although that kind of interaction is incessant, the strength of its action depends on mutual states of interacting DQDs (in-resonance or off-resonance). The logic operations between adjacent DQDs are executed in resonant regime; in the off-resonance the quantum information is stored.

In the proposed register any quantum algorithm could be effectuated via manipulation solely with digital voltage pulses on controlling electrodes that reminds a functioning of an integrated circuit.

The final read-out of the register is performed after decoding into charge states of DQDs and a transmission of current through the wire.

## 1. INTRODUCTION

Universal scalable quantum computers may allow solution of problems with high computational complexity inaccessible for classical computers (Turing machines). So called NP-problems belong to such a class of problems. At the very beginning the quantum computer was devoted to simulation of complex quantum systems, for example, heavy atoms and big molecules including biological ones. Besides, quantum computers can significantly accelerate the processing of big data.

Up to date several companies and institutions have announced that quantum supremacy had been achieved on superconducting or photonic platforms [1-7]. However, whether those calculations could be performed in classical supercomputers with amended algorithms is still open. One thing is clear: the quantum computing demonstrated so far does not yet offer any obvious practical usefulness.

Despite of many alternative approaches to quantum computation (neutral atoms in optical traps, ions in em traps, superconducting qubits, photonic qubits and so on) the general problem is the scalable quantum register including thousands of logical qubits. For error corrections lots of additional physical qubits (ancillas) are required.

On the contrary, we propose to employ an ensemble (or collective) quantum register much more resistant against environment noise including that of phonons and random charges due to obvious averaging the mean state of constituent qubits. In the ensemble register instead of single silicon nanowire, an array of them are placed under common controlling electrodes (gates). It opens up a possibility of large-scale quantum computer fabricated by advanced silicon technology.

Double quantum dots (DQD, two tunnel bound quantum dots with a single electron) as prospective solid-state qubits for quantum computation were put forward in 1999 [8].

A DQD has advantages with respect to a single one. Firstly, two lowest levels (symmetric and antisymmetric states) are quite distant from upper ones. Just those levels serve as basic for a qubit, while the upper levels weakly influence on the qubit evolution. Secondly, for fairly small energy splitting between symmetric and antisymmetric states the decoherence caused by phonons is much suppressed.

It should be outlined that the decoherence caused by phonons is inevitable for solid-state structures, meanwhile, the influence of random charged defects could be, in principle, diminished by advances in technology. However, for the present state of art just charged defects are widely recognized as the primary source of decoherence in solid-state structures.

One more benefit of DQDs is a possibility to control their states and interaction solely with outer electrodes. In Ref. 8 the particular structure was proposed although it looks as quite cumbersome for experimental realization. In Ref. 9 the simplified structure with exchange interaction between quantum dots was put forward. However, this kind of interaction implies the quantum dots to be quite close to each other that complicates metallization and leads to severe cross-talks. Here we put forward a quantum register with Coulomb interaction quite accessible for modern silicon technology.

The pioneering paper [8] has acquired multiple citations. The first experimental realizations were made in Japan [10-13]. The authors used quantum dots defined by electric fields of metallic gates over two-dimensional electron gas in GaAs heterostructure. The investigations have stopped at two DQDs because the structure turned out to be unscalable.

Although the idea of DQD qubit was put forward a long time ago it looks still promising. Recently many research centers over the world draw attention to the realm of quantum computers based on the semiconductor platform, especially on CMOS technology [14-29]. In the review paper [22] the attention is mostly paid to solid-state spin qubits owing to their larger coherence time compared to that of charge qubits. Nevertheless, with respect to the number of operations during a coherence time the charge qubits do not lose to the spin ones.

## 2. DESIGN OF THE QUANTUM REGISTER (INDIVIDUAL AND ENSEMBLE)

In 2011 the quantum register consisting of field-defined DQDs in a silicon nanowire wrapped by dielectric was put forward (Fig. 1) [16]. Quantum algorithms could be effectuated via manipulation solely with digital potential pulses on controlling electrodes (gates) that reminds the operation of a digital integrated circuit. Evidently, this construction is scalable.

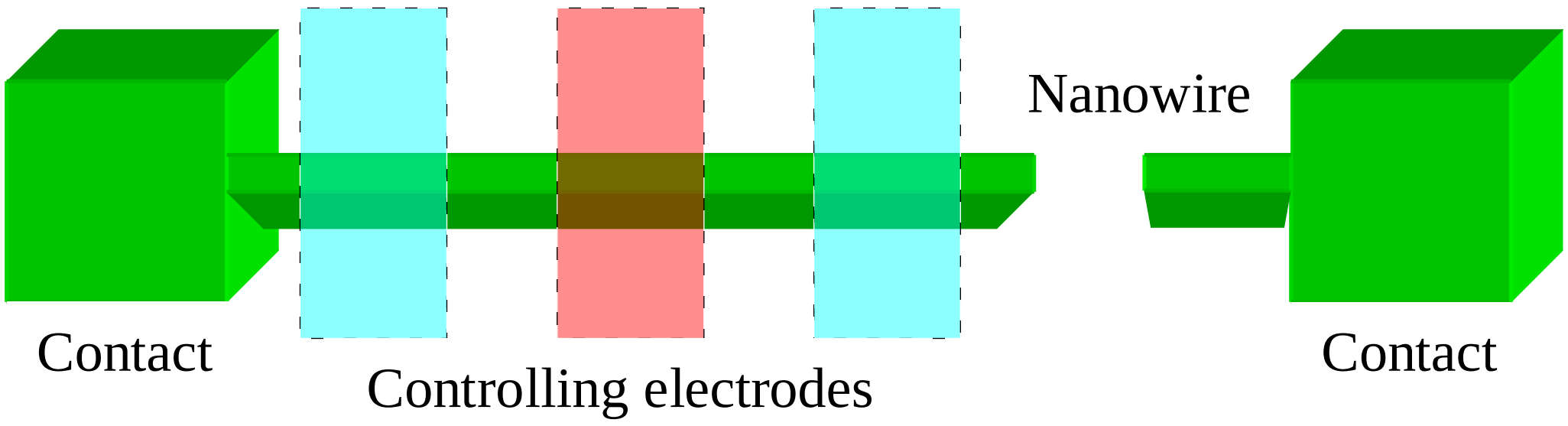

Fig. 1. Schematic view of the *single* quantum register consisting of electrically defined and operated DQDs in an undoped silicon nanowire with controlling electrodes and doped contacts at its ends. The construction reminds that of multi-gate fin-MOSFET transistor [16].

The important advantage of the proposed structure is that all constituent DQDs could be tuned by gate potentials to compensate the impact of stationary random charges which are the everlasting reproach to solid-state charge qubits. The technological variability could be also compensated the same way.

It is readily seen in Fig. 1 that the single register structure could be multiplied. The implementation of the ensemble register consists of a nanowire array with mutual controlling electrodes and contacts (Fig. 2).

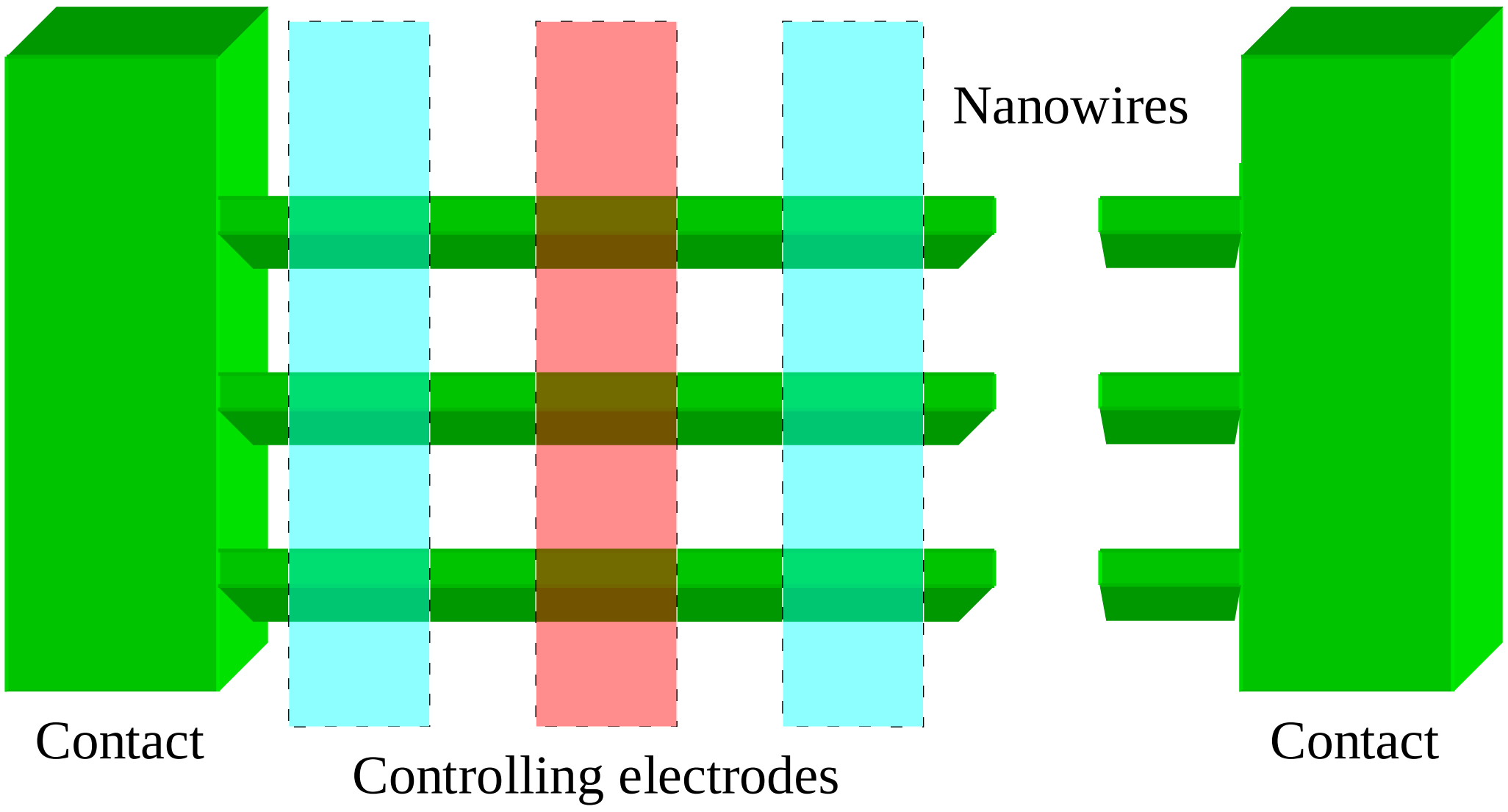

Fig. 2. Schematic view of the *ensemble* quantum resister consisting of electrically defined and operated DQDs in silicon nanowires with mutual controlling electrodes and contacts at the ends.

The ensemble qubits are much more resistant against environment noise including that of phonons and random charges due to averaging and compensation.

One more advantage of the ensemble qubit is a possibility of the macroscopic read-out procedure when the large number of electrons is sequentially kicked out from ensemble quantum dots and the arisen currents are measured.

The schematic cross-sectional view with three controlling electrodes (gates) which electrically define a DQD in a silicon nanowire is depicted in Fig. 3, while the corresponding band diagram with indicated Fermi level in heavily doped contacts is presented in Fig. 4.

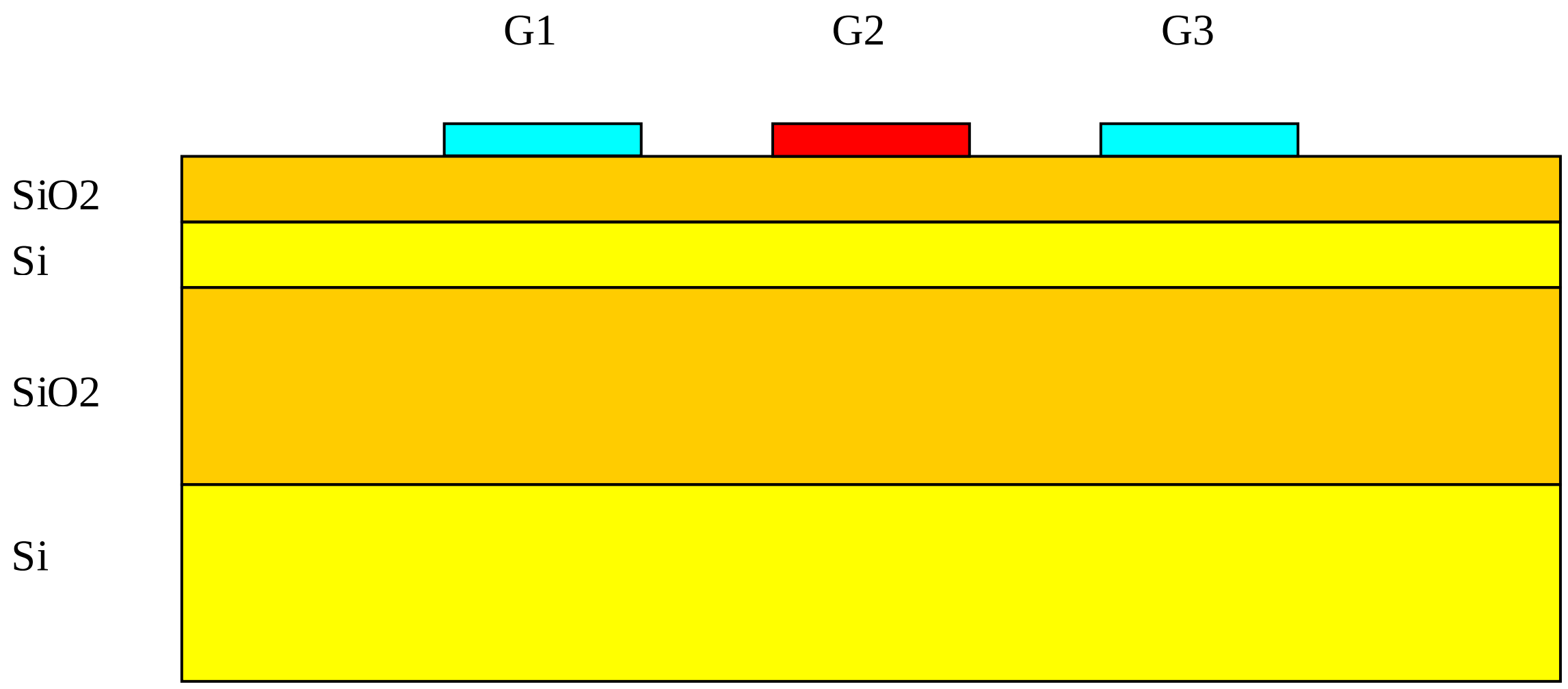

Fig. 3. Schematic cross-sectional view with three controlling electrodes (gates) which electrically define a DQD in a silicon nanowire.

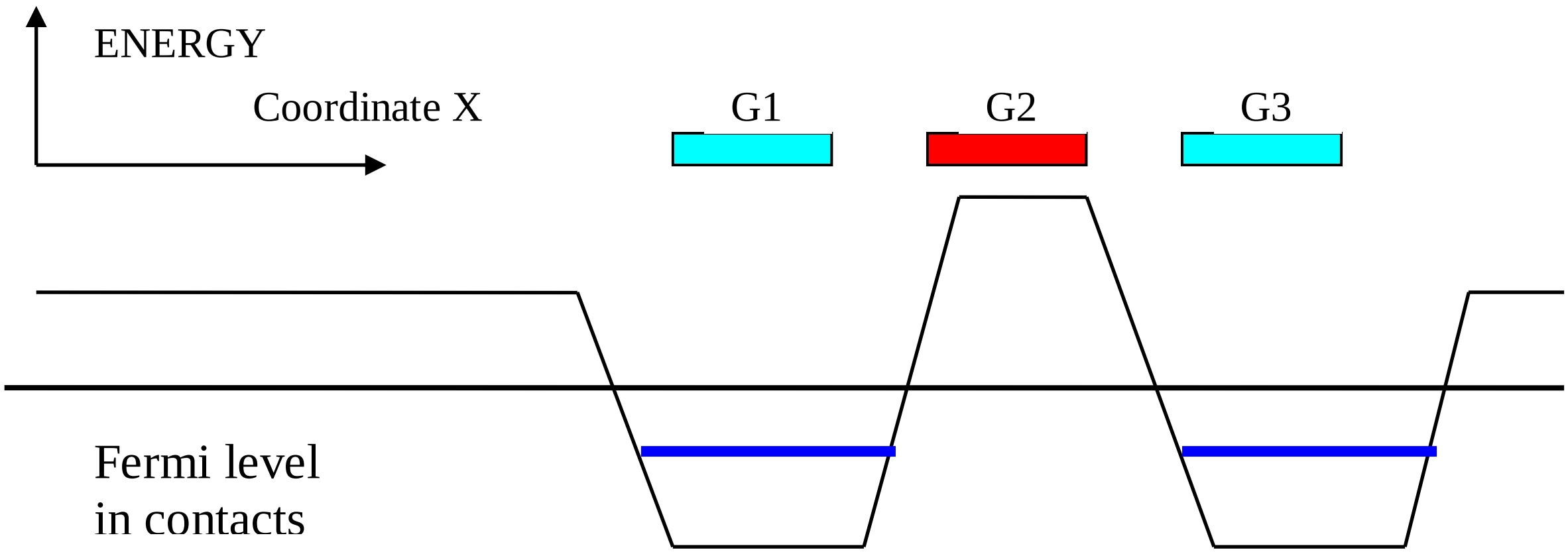

Fig. 4. The longitudinal (along the wire) potential profile (the band bottom) with one double quantum dot. The Fermi level in heavily doped contacts is indicated.

An individual qubit in the proposed register consists of *two* DQDs [9] (Fig. 5) hereafter referred to as 2DQD qubit. The states of constituting DQDs are permanently symmetric state or antisymmetric. Then the quantum information can be encoded and processed without charge transfer between dots. The probability to find electron in a dot is constantly equal to 1/2. This guarantees a linear dynamics of 2DQD qubits necessary for now existing quantum algorithms. Worth noting, the dynamics of qubits with altering charge state is actually nonlinear due to interaction with surrounding dielectrics and metals (polaron effect). All that is a reason to call the discussed 2DQD qubits as 'space qubits' unlike the habitual term 'charge qubits' implying a charge transfer between dots.

One more reference of the 2DQD qubit with respect to the 1DQD qubit is its fairly strong tunneling coupling between dots both in active and idle state which prevents from severe influence of random charges.

The technological advantage of the proposed 2DQD qubit is a relatively large separation between DQDs which simplifies metallization and diminishes cross-talks between controlling electrodes.

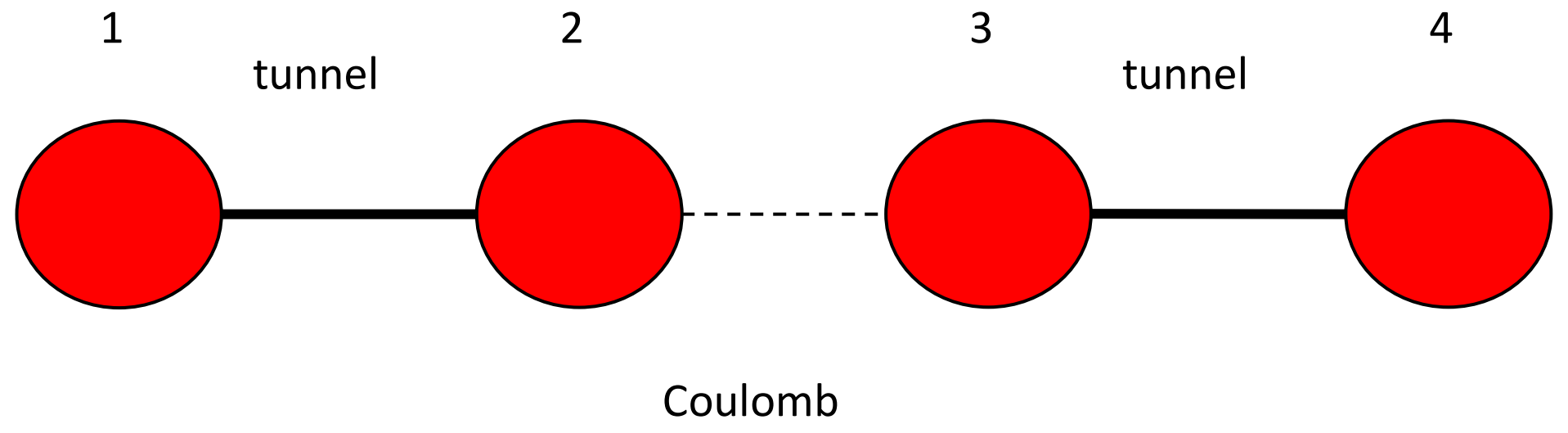

Fig. 5. The schematic view of the proposed 2DQD qubit, It consists of *two* DQDs with tunnel coupling *T* and Coulomb interaction *V* between adjacent dots (between far distant dots the interaction is screened by metallic controlling electrodes).

The computational basis of the 2DQD qubit is discussed below. The final state of the register (read-out) is measured by the current passing through the nanowire between side contacts. Beforehand, a decoding from the space state of a DQD to its charge state should be performed. In the charge state an electron occupies the left or the right dot in a DQD which could be distinguished by the current.

## 3. PERFORMANCE OF THE REGISTER

### 3.1. Initialization

Here we consider the simplest procedure of initialization when an electron is put in one dot of two in a DQD (Figs. 3, 4). The second dot is eliminated by the potential on the corresponding controlling electrode. After that, by manipulating with potentials all DQDs could be put in symmetric $|s\rangle$ or antisymmetric $|a\rangle$ states. This procedure is essentially the Hadamard transformation and it is inverse to decoding performed prior to final readout, which is discussed later.

The electrons in silicon conduction band require thorough consideration accounting for six minima (valleys) with anisotropic effective mass. Therefore, here we deal with heavy holes of the isotropic valence band, their effective mass is $m^*=0.49\ m_0$.

The longitudinal potential relief (the band bottom) at the end of the register is depicted in Fig. 4. The Fermi energy in the contact region is determined by doping level. In the region of undoped nanowire the band bottom is shifted by the energy of transversal quantization

$$E_t \approx \frac{h^2}{4m^* d^2}, \qquad (1)$$

where *d* is the wire diameter, *h* is the Planck constant.

The quantum dots are defined by controlling gates $G_1$ and $G_3$. The barrier height in between is manipulated by the gate $G_2$, while the potential depth is manipulated by the gates $G_1$ and $G_3$. The additional gates near the contacts could be used to block the current through the wire during calculation and unblock it during the final read-out.

For the dot to be occupied the ground level in a dot should be fairly below the Fermi level in contacts:

$$\Delta E_0 > kT, \qquad (2)$$

where *kT* is the temperature in energy units (*k* is the Boltzmann constant), $\Delta E_0$ is the ionization energy of a quantum dot.

The second electron is repelled when the Coulomb energy of two electrons in a dot is sufficiently high:

$$E_C \approx \frac{e^2}{\kappa l}\ln\frac{l}{d} > \Delta E_0, \qquad (3)$$

where $l$ is the length along the wire of the electrode $G_1$ and $\kappa$ is the permittivity. For the parameters: d = 5 nm, l = 10 nm, and $\kappa_{Si}$ =12, the Coulomb energy could be roughly estimated as $E_C$=0.01 eV=100 K.

Actually, the weak inequalities (2) and (3) should be replaced by strong ones for all dots in a large register to be occupied only by one electron (hole) for sure. The exact conditions of perfect initialization could be revealed only via thorough simulation. In particular, one must take into account the density of states in continuous spectrum. However, according to strong exponential dependence of occupation probability on energy at low temperature, the perfect initialization of the register with large number of DQDs at T=4K seems plausible.

## 3.2. The description based on four-point two-particle wave functions

It is crucial that we employ the two-particle wave functions [30]. All descriptions based upon mean Coulomb potential (Hartree's approach) calculated via $|\psi|^2$ turned out to be deficient.

For proper functioning of the register the hierarchy must be observed: V << T << $\varepsilon_0$, where $\varepsilon_0$ is the ionization energy of a quantum dot, T is the tunnel coupling between dots, and V is the energy of Coulomb interaction between dots. That provides a fairly strong and equal confinement of an electron in dots for $|s\rangle$ and $|a\rangle$ states and constant Coulomb interaction between DQDs regardless of their states. After that one can proceed to the reduced description based on four-point two-particle wave functions.

According to quantum mechanics the two-particle wave function for four dots is described by four complex amplitudes $\psi_{13}$, $\psi_{23}$, $\psi_{14}$, $\psi_{24}$. In the system under consideration the first electron occupies dots $i = 1, 2$, and the second dots $j = 3, 4$. The value $|\psi_{ij}|^2$ is a probability to find one electron in the dot $i$, while the second electron in the dot $j$. In particular, probability to find an electron in the dot $i$ regardless of the position of the second electron is $|\psi_{i3}|^2 + |\psi_{i4}|^2$. For our particular qubit (Fig. 5) the equations are:

$$\begin{cases} i\hbar\dfrac{\partial\psi_{13}}{\partial t} = -T_1\psi_{23} - T_2\psi_{14}, \\ i\hbar\dfrac{\partial\psi_{23}}{\partial t} = -T_1\psi_{13} - T_2\psi_{24} + V\psi_{23}, \\ i\hbar\dfrac{\partial\psi_{14}}{\partial t} = -T_1\psi_{24} - T_2\psi_{13}, \\ i\hbar\dfrac{\partial\psi_{24}}{\partial t} = -T_1\psi_{14} - T_2\psi_{23}, \end{cases} \qquad (4)$$

where $T_1$>0 is an energy of tunnel coupling in the left DQD, $T_2$>0 is an energy of tunnel coupling in the right DQD, $V$ is an energy of Coulomb interaction between adjacent dots 2 and 3. The system of equations is *definitely linear*.

The tunnel couplings $T_1$ and $T_2$ could be switched on ($T$>0) or switched off (T=0). Unfortunately, one cannot do the same with the Coulomb interaction $V$ which is permanent. Luckily, although the Coulomb interaction is unceasing the strength of its action on qubits depends on mutual qubit states.

## 3.3. Manipulations with 2DQD qubit

For the individual 2DQD qubit two basis states are sufficient: $|sa\rangle$ (the left DQD is in a symmetric state while the right one is in antisymmetric state) and $|as\rangle$ (vice versa). These states arise after the initialization. In general, the superpositional states in this basis are entangled. They all belong to the Hilbert subspace with a permanent filling factor of 1/2 at every quantum dot. This 2DQD qubit behaves as a usual pseudo-spin qubit controlled by voltage pulses.

However, the states $|ss\rangle$ and $|aa\rangle$ can also arise during the execution of two-qubit operations. Therefore, the 2DQD qubit basis and the corresponding Hilbert subspace must be extended. Nevertheless, the additional states appear only in combinations which still belong to the Hilbert subspace with a dot filling factor permanently equal to 1/2.

Of all the one-qubit operations, here we consider only the NOT gate, the implementation of which is quite straightforward. This gate means the transitions $|s\rangle \to |a\rangle$ and $|a\rangle \to |s\rangle$ in a DQD. For that two voltage pulses are simultaneously used. One pulse applied to the central electrode switches off the tunneling (T=0), and a voltage $\pi$-pulse applied to the left or right electrode alters the phase. For the execution of the $\sqrt{NOT}$ gate the half-duration pulses are required.

Other operations like phase and amplitude shifts, and the Hadamard transform in the extended basis are thoroughly discussed in Appendix of Ref. 9.

## 3.4. SWAP$_1$ operation between two DQDs

The SWAP$_1$ operation between adjacent DQDs is crucial in the quantum register under consideration. It is used for an amplitude shift of a particular 2DQD qubit as well as the basic operation between two 2DQD qubits. When two DQDs possess an equal separation between symmetric and antisymmetric states (tunnel coupling) the SWAP operation can be efficiently carried out due to entanglement resonance.

This operation also allows the transportation of a qubit state along the register and organization of interaction between any qubits in the chain. Otherwise; the interaction mediated by photons could be employed between distant QDs [31-33].

Besides, the √SWAP operation is fulfilled by the voltage pulse of a half-duration. This operation  is universal along with CNOT.

When the tunneling splitting in DQDs much differs, the Coulomb interaction weakly influences on their states that enables retention (idle state).

At the very beginning, it is convenient to change variables: $\tilde{t} = Tt/\hbar$ и $\tilde{V} = V/T$. After deleting the tildes the system of equations (4) for two identical DQDs with equal tunneling coupling $T_1=T_2=T=1$ looks like

$$\begin{cases} i\dfrac{\partial\psi_{13}}{\partial t} = -\psi_{23} - \psi_{14}, \\ i\dfrac{\partial\psi_{23}}{\partial t} = -\psi_{13} - \psi_{24} + V\psi_{23}, \\ i\dfrac{\partial\psi_{14}}{\partial t} = -\psi_{24} - \psi_{13}, \\ i\dfrac{\partial\psi_{24}}{\partial t} = -\psi_{14} - \psi_{23}. \end{cases} \qquad (5)$$

Here one electron occupies dots 1 and 2 while another occupies dots 3 and 4. An indistinguishability of particles is omitted as it does not affect the result.

The evolution could be precisely described when the eigenfunctions are derived. There are four of them. In general, one should search for the solutions of the kind $\psi_{ij} \sim exp(-i\varepsilon t)$.

$$\begin{cases} \varepsilon\psi_{13} = -\psi_{23} - \psi_{14}, \\ \varepsilon\psi_{23} = -\psi_{13} - \psi_{24} + V\psi_{23}, \\ \varepsilon\psi_{14} = -\psi_{24} - \psi_{13}, \\ \varepsilon\psi_{24} = -\psi_{14} - \psi_{23}. \end{cases} \quad (6)$$

The first and the last equations coincide. One of them could be replaced by the normalization (the eigenfunctions are real):

$$\psi^2_{13} + \psi^2_{14} + \psi^2_{23} + \psi^2_{24} = 1 \quad (7)$$

Two eigenfunctions $\psi_1$ and $\psi_2$ for the weak Coulomb interaction (V << 1) are close to |ss> and |aa>, respectfully. They could be easily deduced with the help of perturbation approach. It follows that the energy of the states is shifted and becomes equal to $\varepsilon_1 = -2 + V/4$ for $\psi_1$ and $\varepsilon_2 = +2 + V/4$ for $\psi_2$. The correspondent wave-function are slightly corrected by the value of the order of V which could be ignored if V << 1. The action of the Coulomb interaction on states |sa> and |as> is *strong* whatever is the value of V. In absence of Coulomb interaction (V=0) those states are degenerate, namely, they have the same energy equal to zero. The Coulomb interaction, whatever small it is, lifts the degeneracy. Then, in the case, the system (6) has two simple *exact* solutions:

$$\psi_3 = [\psi_{13}, \psi_{14}, \psi_{23}, \psi_{24}] = [1,\ 0,\ 0,\ \text{-}1] / \sqrt{2} = \left(|sa\rangle + |as\rangle\right) / \sqrt{2} \quad (8)$$

with the energy $\varepsilon_3 = 0$, and

$$\psi_4 = [\psi_{13}, \psi_{14}, \psi_{23}, \psi_{24}] = [0,\ -1,\ 1,\ 0] / \sqrt{2}\ exp\left(-iVt\right) = \left(|sa\rangle - |as\rangle\right) / \sqrt{2}\ exp\left(-iVt\right) \quad (9)$$

with the energy $\varepsilon_4 = V$.

Those states are *significantly* quantum in sense of strong quantum correlation. It is readily seen that the probability to find an electron in any dot exactly equals 1/2. At the same moment, there is a strong correlation between nearby dots 2 and 3. For the state $|\psi_3\rangle$: if one detects the first electron in the dot 2, the wave-function of the second electron collapses to the distant dot 4. Therefore, the Coulomb energy in this state equals zero. On the contrary, for the state $|\psi_4\rangle$: if one detects the first electron in the dot 2, the wave-function of the second electron collapses to the nearby dot 3. Therefore, the Coulomb energy in this state equals V.

Let us consider the $SWAP_1$ operation starting from the initial state $\psi(0) = |sa\rangle$. Its evolution stems from the equations (8) and (9):

$$\psi(t) = \frac{1}{\sqrt{2}}(\psi_3 + \psi_4) = \frac{1}{2}[|sa\rangle(1 + e^{-iVt}) + |as\rangle(1 - e^{-iVt})] \quad (10)$$

Hence, in the lapse of time π/2V the $sqrtSWAP_1$ occurs, in the lapse of time π/V the $SWAP_1$ (exchange of states) is fulfilled, in the lapse of time 2π/V the system returns to the initial state, and so on. The evolution of the initial state $\psi(0) = |as\rangle$ is analogous.

As according to the eigenfunctions $\psi_1$ and $\psi_2$ they are almost intact with respect to Coulomb interaction, eventfully, it means that the $SWAP_1$ operation is fulfilled in extended 2DQD basis (including $|ss\rangle$ and $|aa\rangle$ states).

Worth noting, the $SWAP_1$ operation occurs when the digital pulse applied to central electrode in a DQD brings both DQDs in resonance (the energy separations between symmetric and antisymmetric states coincide). When the pulse ceases the operation is aborted in the off-resonance regime (the energy separations between symmetric and antisymmetric states fairly differ).

## 3.5. $SWAP_2$ operation between two 2DQD qubits

The basis of the Hilbert sub-space for operations between 2DQD qubits is [9]:

$$\begin{array}{lcl} \textit{Left qubit} & & \textit{Right qubit} \\ |sa\rangle & _________ & |sa\rangle \\ |ss\rangle & _________ & |aa\rangle \\ |sa\rangle & _________ & |as\rangle \\ |as\rangle & _________ & |sa\rangle \\ |aa\rangle & _________ & |ss\rangle \\ |as\rangle & _________ & |as\rangle \end{array} \tag{11}$$

All these states have *zero* energy (the Coulomb interaction is excluded), and the probability of finding the electron in either of the quantum dots remains equal to 1/2.

The foregoing $SWAP_1$ operation between two DQDs makes it possible the $SWAP_2$ operation between 2DQD qubits. Its circuit presentation is depicted in Fig. 6

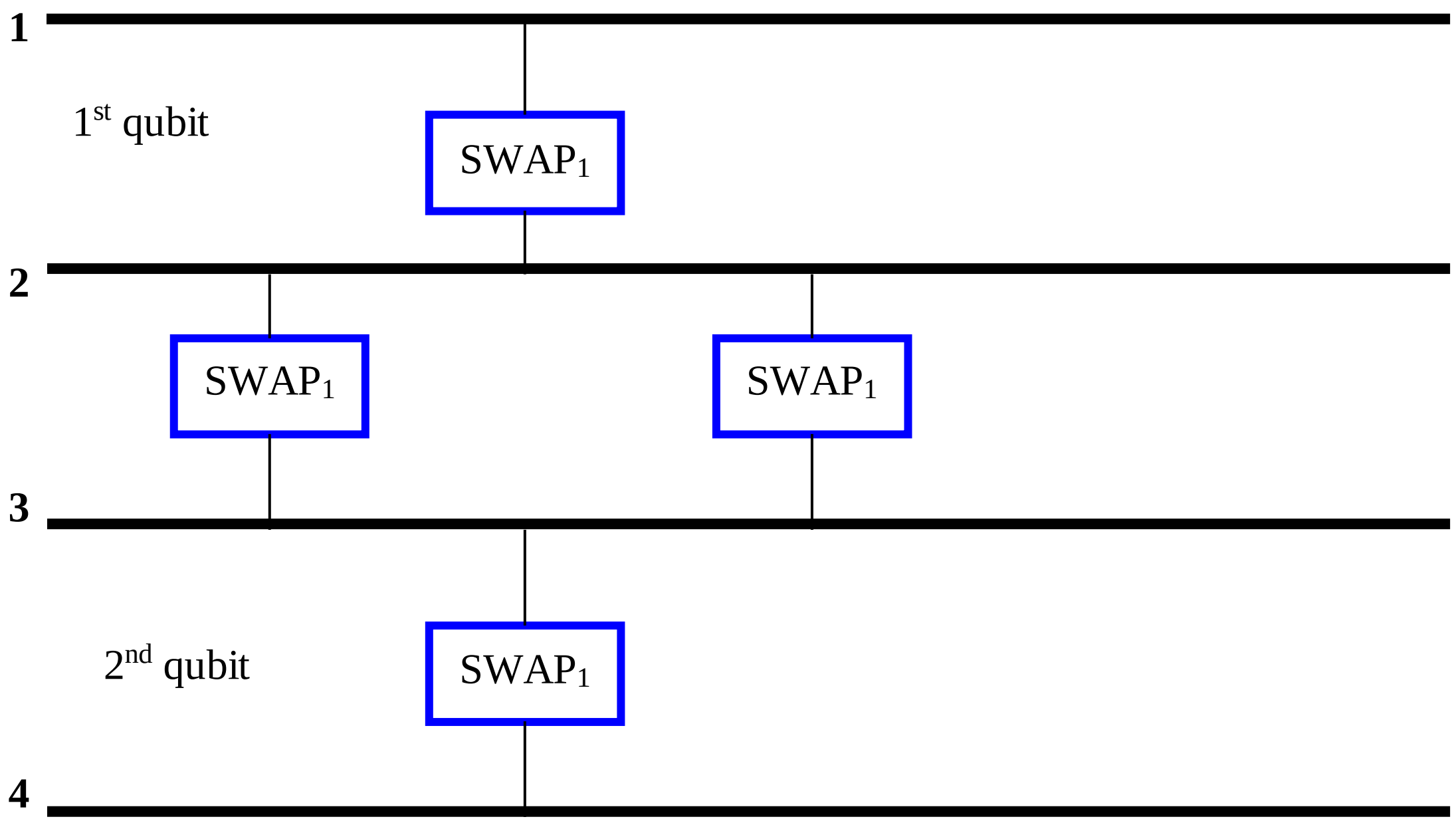

Fig. 6. Circuit presentation of $SWAP_2$ operation between 2DQD qubits. When repeated twice it results in the identity operation *I*. The first qubit consists of 1st and 2nd DQDs, the second qubit consists of 3d and 4th DQDs. Here “$SWAP_1$” denotes the operation between two adjacent DQDs.

It is important to note that when performing a $SWAP_1$ operation between specific DQDs in the resonant regime, all other pairs must be off-resonant at the moment.

Analogously, the circuit presentation of $\sqrt{SWAP_2}$ gate is portrayed in Fig. 7.

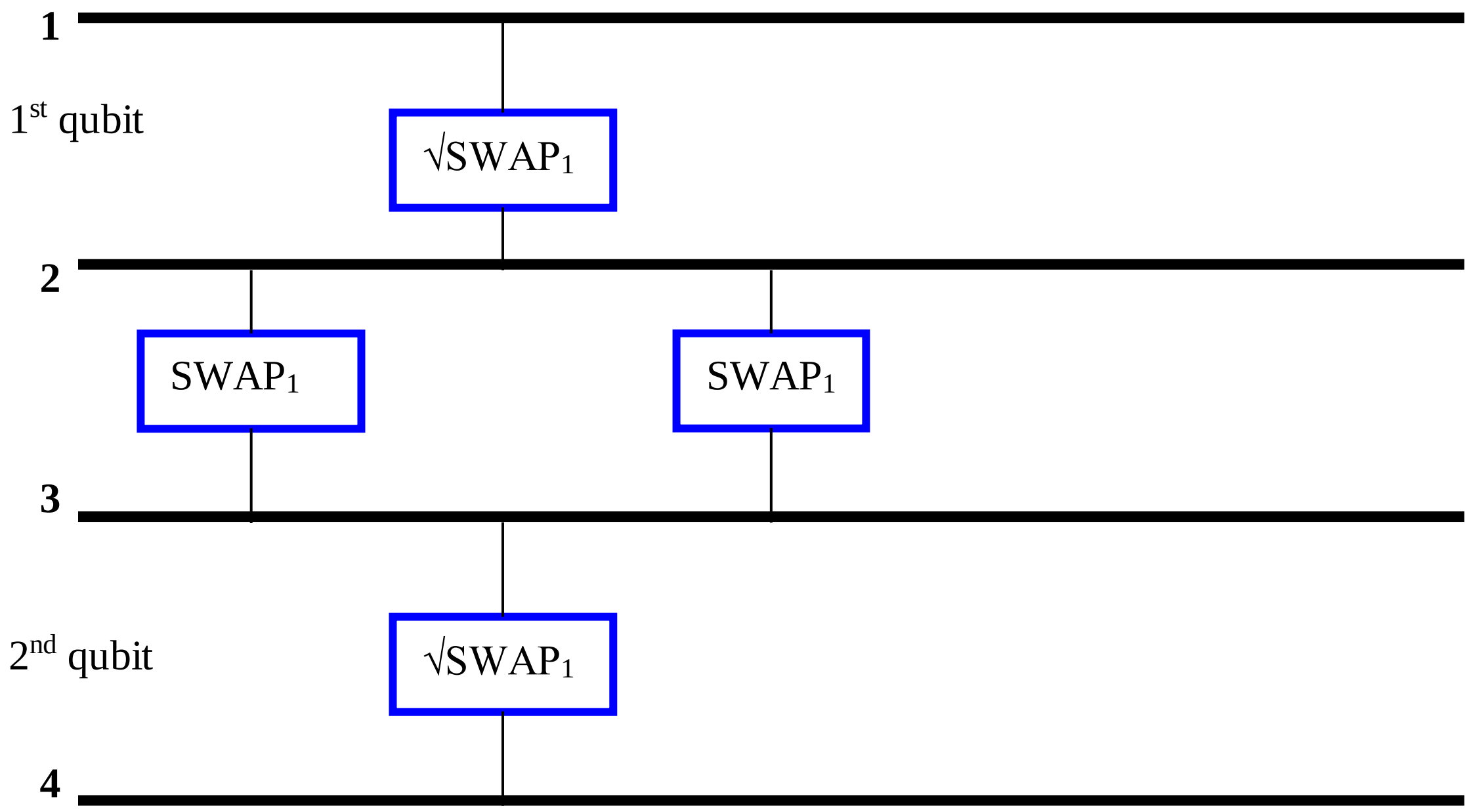

Fig. 7. Circuit presentation of √$SWAP_2$ operation between 2DQD qubits. When repeated twice it results in SWAP operation in Fig. 6. Here "$SWAP_1$" and "√$SWAP_1$"denote the operations between two adjacent DQDs.

It should be outlined that two states from the basis (11) appear only in combination $|ss\rangle|aa\rangle + |aa\rangle|ss\rangle$, corresponding to zero energy and filling factor of a dot equal to 1/2.

The extended basis (11) makes the realization of the CNOT gate from the √$SWAP_2$ gate non-trivial but still possible [9]. However, the √$SWAP_2$ gate is itself universal.

## 3.6. Retention in the idle state

It looks like the state of qubits could be "frozen" when the tunnel coupling inside DQDs is switched off (T=0). Unfortunately, it is not fully the case. Although the state with tunneling disabled is highly robust against phonon relaxation, it is quite vulnerable to the influence of random charges capable of easily altering the qubit's phase.

Luckily, there is another possibility to save information in an idle state (retention). When tunnel coupling in two adjacent DQDs differs and obey the conditions $T_1,\ T_2,\ |T_1 - T_2| >> V$ (off-resonant regime) the states of DQDs are almost intact to Coulomb interaction.

## 3.7. Read-out

Long time ago, the method to perform the read-out of a quantum dot charge and spin state with the help of current in the regime of Coulomb or spin blockade was put forward in the Ref. 34 and then developed in [35]. Here we scrutinize this fruitful idea.

To read out the information accumulated in the resultant register in binary code one should measure the states of DQDs, that is, to distinguish the symmetric state of a DQD from the antisymmetric one.

Therefore, the measurement of the final state of the register is performed in two steps. Firstly, the symmetric and antisymmetric states of DQDs are decoded by manipulations with gate potentials into charge states when an electron is placed into left or right dot [9]. Here we demonstrate this procedure carefully.

After an additional potential G is applied to the right dot the equations for the wave-function ($\psi_L$, $\psi_R$) turn into

$$\begin{cases} i\dfrac{\partial\psi_L}{\partial t} = -\psi_R, \\ i\dfrac{\partial\psi_R}{\partial t} = -\psi_L + G\psi_R. \end{cases} \qquad (12)$$

The most interesting is the case with $G$= - 2. The exact solution argues that in the time lapse equal to $t$=π/2√2 the electron from the $|a\rangle$ state entirely transfers to the left dot, and from the $|s\rangle$ state it entirely transfers to the right dot, that is, the decoding is fulfilled. In fact, this is the Hadamard transformation of the basic states $|a\rangle$ and $|s\rangle$. The diagram in Fig. 5 explains this process. The essence is that the transitions occur between states with equal mean energies.

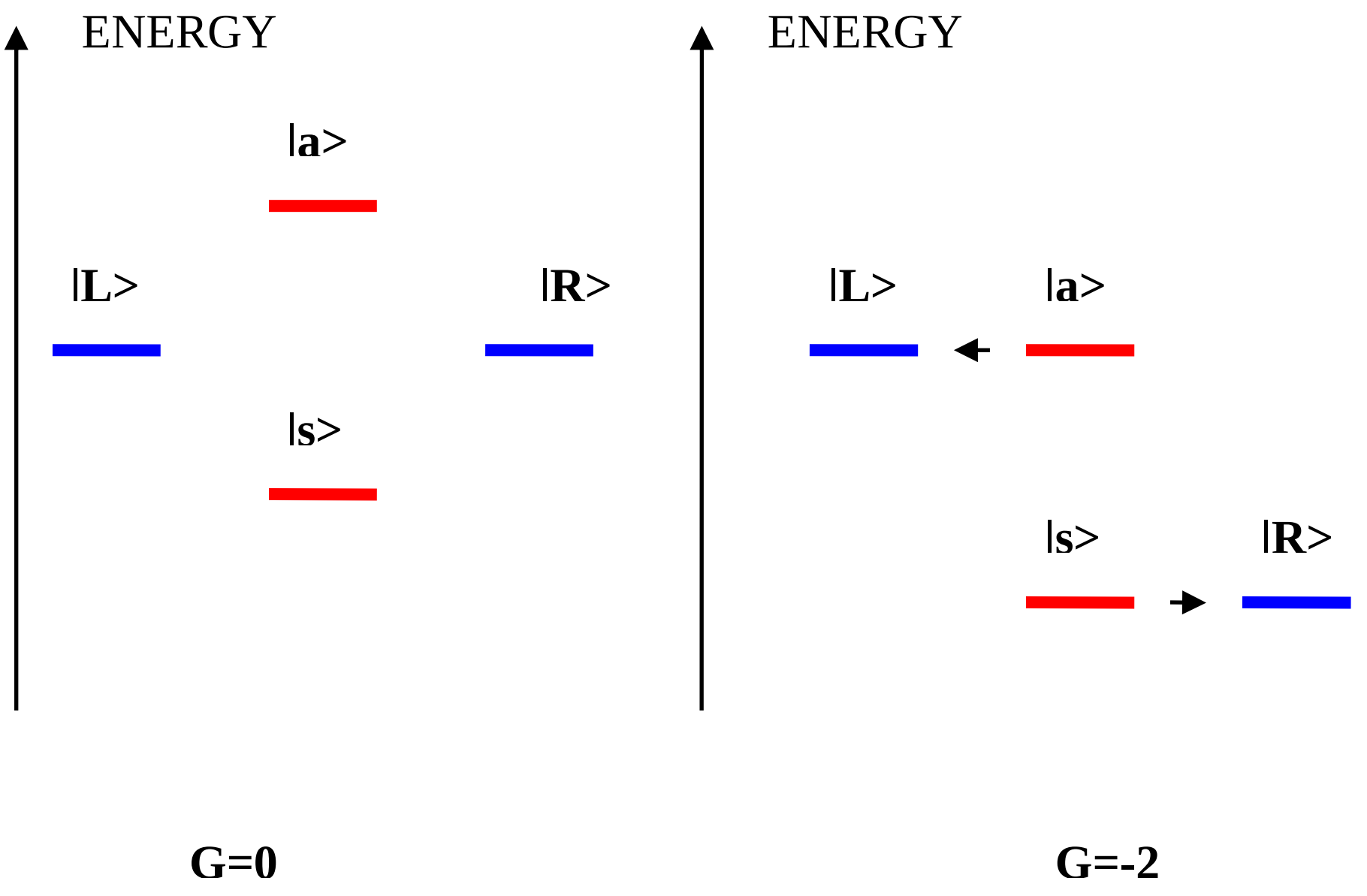

Fig. 8. The mean energy diagrams: left – no voltage applied (G=0), right – the voltage is applied to the right dot (G= -2).

After decoding the measurement should reveal what dot of two in the DQD is occupied. It is performed with the help of weak current passing through the nanowire in the regime of Coulomb blockade of current [34, 35]. Worth noting, during calculations there are no free electrons in the wire.

The Coulomb blockade can be managed as follows. One dot in the register is chosen as a target when all other dots are "deepened" by the potential of corresponding electrodes (Fig. 9). If the target dot is filled there is a potential hump for electron passing through the wire. If the target dot is empty there is no potential hump and the transmission coefficient is larger (Fig. 10). To readout the register one should measure 1/4 dots sequentially, one by one.

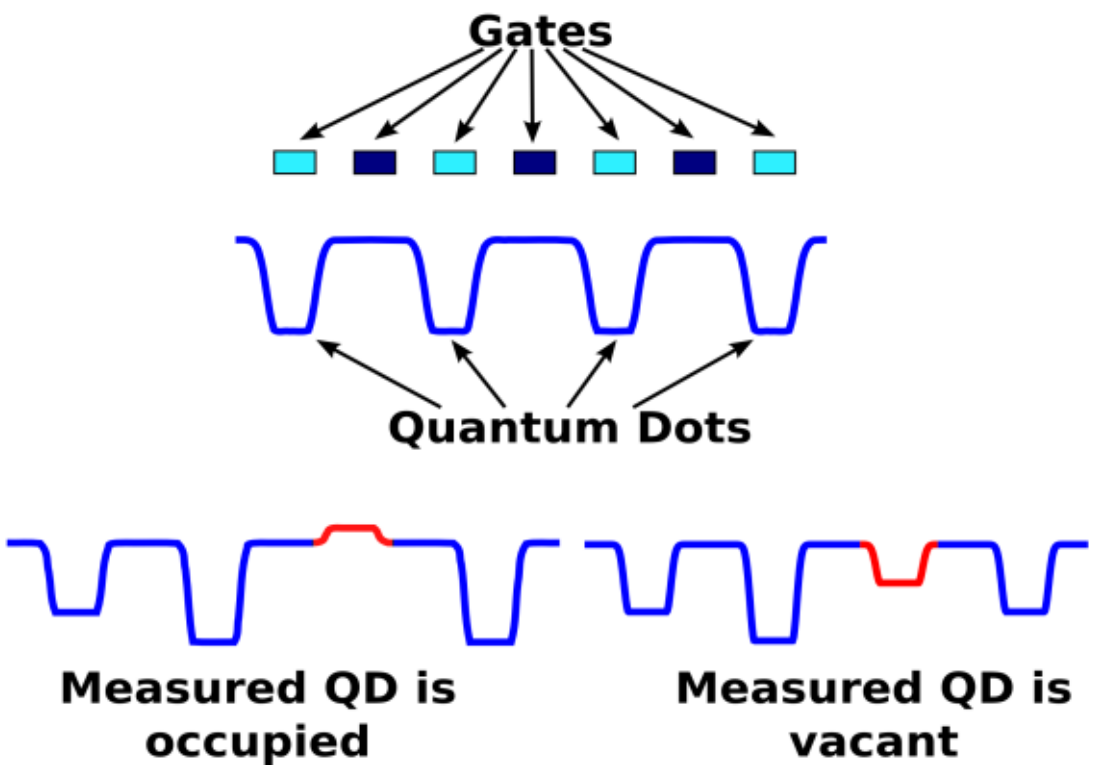

Fig. 9. Potential relief confining electrons and created prior to the measurement of a target quantum dot: there is a hump for occupied dot and a pit for unoccupied one. All other dots are "deepened" [35].

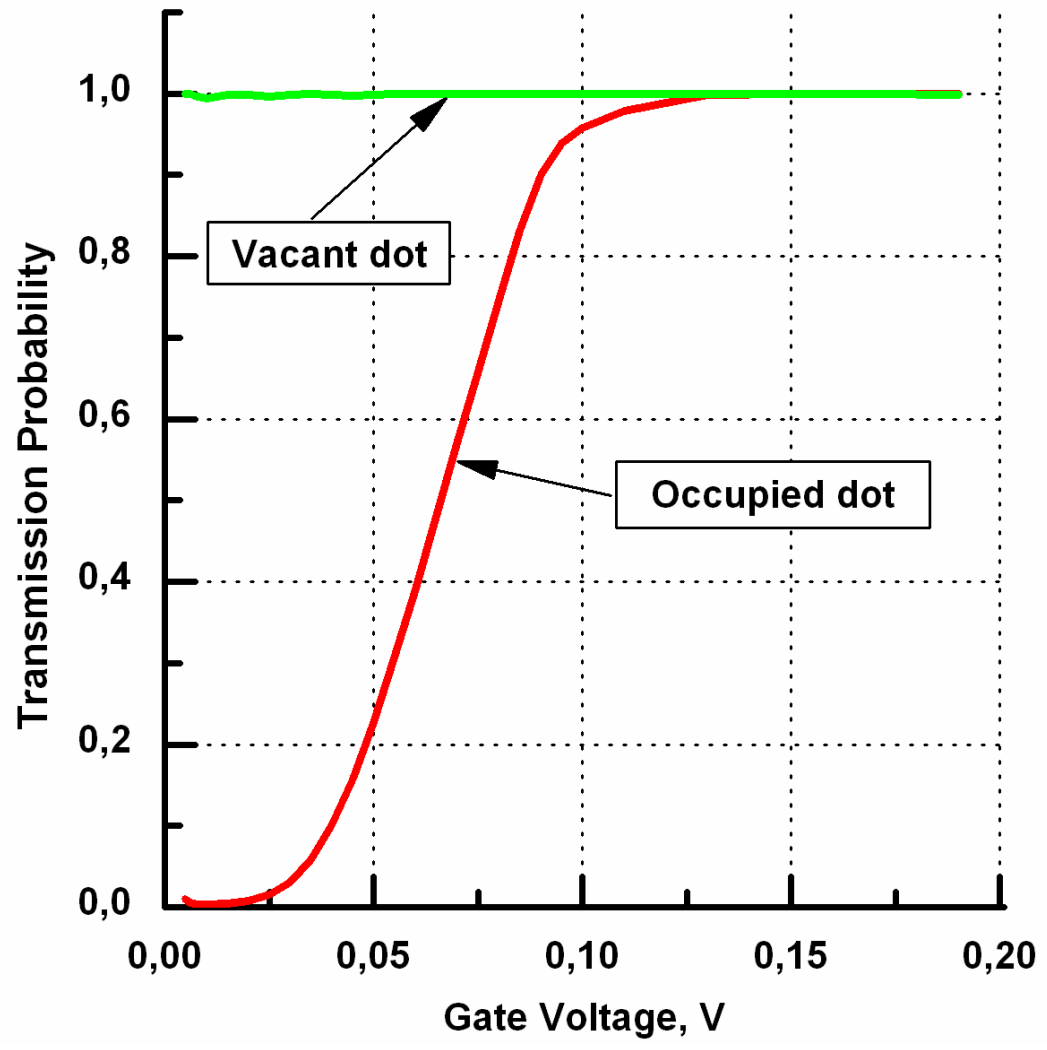

Fig. 10. Mean transmission coefficient through the channel vs. gate voltage for vacant and occupied dots. The Fermi energy of incident electrons is 25 meV and ambient temperature is 4K [35].

Really, the number of dots to be immediately measured in this way is restricted by trapping of free electrons in the wire. For the sake of not overload the paper, here we provide only a qualitative argument for trapping probability to be small. Indeed, the matrix element of transition of a free electron to a quite deep state in a dot with emission of an acoustical phonon is very small. Whatever it is, a computation could be repeated and another group of dots measured. It should be recalled that the ensemble register allows a macroscopic measurement in the unique cycle.

## 4. SIMULATION OF THE STRUCTURE WITH A SINGLE ELECTRON

The problem of simulation of solid-state structures with a single electron is thoroughly discussed in Ref. 36. Here we touch on it only schematically.

The milestone of the correct simulation is to exclude the action of an electron upon itself. Only the field created by the environment in response to that of an electron should be involved into the evolution equation. There are two different approaches to cope with the problem.

The first strategy is based on the response of environment to the fictitious charge density $e|\psi|^2$ ($\psi$ is a wave-function) [36-44]. The simulation is fulfilled in two steps. Firstly, the Poisson equation is solved with the charge density $e|\psi|^2$ emerged in dielectrics and surrounded by metal electrodes. Secondly, the Poisson equation is solved with the charge density $e|\psi|^2$ in empty space (vacuum). The difference of acquired potentials is substituted to the Schrödinger equation to obtain an updated wave function, and so on. In the case, the Schrödinger equation is apparently non-linear. By the way, this nonlinearity is a fundamental of the theory of polaron [37, 38]. In general, the severe consequence for the qubits based on single or double quantum dots with an altering charge state appears: they indispensably become nonlinear.

At least, for weak nonlinearity it means a restriction of accuracy of quantum operations for conventional quantum computing based on linear Hamiltonians. Intriguingly, nonlinear quantum mechanics implies polynomial-time solution for NP-complete and #P problems [45].

For strong nonlinearity the effect may become much more pronounced. For instance, if a DQD is elongated near a metal surface the breaking of symmetry originates when the Coulomb interaction with metal exceeds its tunnel coupling. The ground state becomes twice degenerate (an electron primarily occupies the left dot or the right one instead of the initial symmetric state with equal occupation probability). Our qubit constituting of two DQDs remains linear as its charge state does not change during computation.

Evidently, this approach is valid if only the frequency of oscillations in the quantum system is small compared to the frequencies of relaxation in ambient metal and dielectrics. In the proposed register all quantum operations are sufficiently slow to meet this requirement.

The second strategy is based on the response of environment to a point charge [36 and refs herein]. The corresponding potential energy is substituted to the Schrödinger equation which remains linear. This approach completely denies an existence of polarons. Moreover, after the wave function is calculated a reasonable question arises: what is a polarization state of environment, in particular, that of metal electrodes. The answer does not exist, and this is the main drawback of the second strategy. Therefore, in the present consideration we adhere to the first strategy.

It matters that the previously excluded self-field of an electron should be involved into calculation of interaction between quantum dots. For fairly large distance (which is relevant to the quantum register under consideration) this field could be approximated with that of a dipole as quantum dots are capped by metal electrodes.

## 5. SEVERAL ESTIMATIONS FOR REALISTIC STRUCTURES

Here we make several estimations of the values used in foregoing equations for realistic structures which could be manufactured by modern silicon technology.

The parameters are as follows: the silicon wire diameter is 5 nm, the gate length $l$ along the wire and gate separation in a DQD corresponds to the technology node 10 nm, the distance between DQDs equals L=200 nm, the thickness of dielectric under gate equals $d_{SiO2}$=5 nm, the permittivities are $\kappa_{Si}$=12, $\kappa_{SiO2}$=4 for silicon and silicon oxide, respectively.

It should be highlighted that the large distance L between DQDs is beneficial for metallization technology and suppression of cross-talks among electrodes. At the same moment, the larger is a distance the more probable is appearance of a random charge.

The energy of dipole-dipole interaction between adjacent DQDs equals

$$V = \frac{e^2}{\kappa_{SiO2}} \frac{1}{L} \frac{4d_{SiO2}^2}{L^2} . \qquad (13)$$

The electric dipoles originate because all dots are capped by metallic gates. For above parameters one gets V= $10^{-6}$ eV that corresponds to the frequency 1 ΓHz. Just this value determines the time of the major SWAP operation in the register. It is evident that this value can be increased or decreased in accordance with formula (13) to align with the capabilities of the control elements.

The tunneling coupling in a DQD varies from practically zero (for fairly high potential barrier between dots) up to its maximal value achieved when the barrier is withdrawn. In the case the value $T_{max}$ is merely the energy gap between symmetric and antisymmetric states in a DQD without a barrier:

$$T_{\max} = \frac{3h^2}{8m^* D^2}, \qquad (14)$$

where $h$ is the Planck constant, D= 50 nm is the distance between quantum dots, $m^*$ is the effective mass. For $m^*=m_0$ one obtains $T_{max}= 5\ 10^{-22}$ J= 3 meV. The ratio $V/T_{max}$ could be evaluated as 2.5 $10^{-4}$. This means that the SWAP operation between equal states ($|ss\rangle$ or $|aa\rangle$) slightly disturbs them.

## 6. DECOHERENCE

The indispensable source of decoherence in a solid-state system is an interaction with acoustical phonons of various kinds [46]. At any rate, this type of decoherence could be suppressed at low temperatures.

Here we mostly pay attention to the harmful source caused by random charged defects. At least, this is the everlasting reproach to all solid-state quantum computer structures.

In the register under consideration all quantum dots are capped by metallic electrodes (gates). Therefore, the action of outer charges is screened. Moreover, the energy shift does not affect the SWAP operation because it occurs in the resonant regime when tunnel splitting in DQDs coincides. The latter is defined solely by tunnel gate voltage.

In general, the action of a random charge becomes severe when it exceeds tunnel splitting in a DQD. Therefore, the preference of the 2DQD qubit with respect to the 1DQD qubit is its fairly strong tunneling coupling between dots both in active and idle state which prevents from severe influence of random charges.

Evidently, a random charge substantially deteriorates a DQD functioning when it is placed inside. Nowadays, the silicon technology looks as the most perfect. Indeed, the concentration of bulk defects (residual dopants) in industrial silicon wafers equals $10^{15}$ $cm^{-3}$. The probability to find a random charge in a DQD with dimensions 10x10x10 $nm^3$ equals $10^{-3}$. The $Si/SiO_2$ interface possesses the record surface concentration of defects among all semiconductor/dielectric interfaces, i.e. $10^{10}$ $cm^{-2}$ [47]. The probability to find a random charge in the DQD still equals 0.1. Therefore, a struggle against this kind of defects is challenging.

Worth noting, the action of constant (built-in) charges could be eliminated by tuning the gate voltages. Besides, the ensemble qubit seems as a promising solution as well.

## CONCLUSION

An implementation of a solid-state quantum register based on electron space states in field-defined double quantum dots (DQD possesses one electron in two adjacent tunnel bound dots) in an ultrathin silicon nanowire is discussed. To some extent, the structure reminds that of a field-effect transistor with multiple controlling electrodes (gates). Scalability is audible and it opens up a possibility of large-scale quantum computer fabricated by advanced silicon technology. All DQDs could be tuned by gate potentials to compensate the impact of stationary random charges which are the everlasting reproach to solid-state charge qubits.

Moreover, the structure of silicon 2DQD qubits could be developed into an ensemble quantum register where instead of single silicon nanowire an array of them (1000 is quite easy to fabricate nowadays) are placed under common controlling electrodes. Those ensemble qubits are much more resistant against environment noise including that of phonons and random charges due to averaging and compensation. One more advantage of the ensemble qubit is a possibility of the macroscopic read-out procedure when the electrons are sequentially kicked out from ensemble quantum dots and the arisen current is measured.

Quantum algorithms could be effectuated via manipulation solely with digital potential pulses that reminds an operation of an integrated circuit.

It is crucial that an individual qubit consists of *two* DQDs. The quantum information is encoded and processed inside the Hilbert subspace without charge transfer between dots. The probability to find an electron in a dot constantly equals 1/2. This guarantees a linear dynamics of qubits necessary for now existing quantum algorithms. Worth noting, the dynamics of qubits with altering charge state is more or less nonlinear due to interaction with surrounding dielectrics and metals (polaron effect). Even for weak interaction the accuracy of quantum operations becomes limited.

One more preference of the 2DQD qubit with respect to the 1DQD qubit is its fairly strong tunneling coupling between dots both in active and idle state which prevents from severe influence of random charges.

The basic two-qubit operations in the system are SWAP and √SWAP. The latter operation is universal as well as CNOT. The two-qubit operations are performed by Coulomb interaction. Although that kind of interaction is incessant, the strength of its action depends on mutual states of interacting DQDs (in-resonance or off-resonance). The resonant regime means an equal tunnel splitting in interacting DQDs when a quantum operation is performed. In the off-resonant regime when the tunnel splitting in interacting DQDs differs the quantum information is stored (retention or idle state).

The final read-out of the register is performed after decoding into charge states of DQDs, in fact, this procedure is the Hadamard transform. Then a transmission of current through the wire in a regime of Coulomb blockade reveals whether a particular quantum dot is occupied or not. The spin blockade of current enables the measurement of spin state of a dot. For the ensemble qubits the macroscopic measurement predominates.

The technological advantage of the proposed 2DQD qubit is a relatively large separation between DQDs which simplifies metallization and diminishes cross-talks between controlling electrodes. Indeed, the DQDs constituting the register could be manufactured on the base of sub10-nm silicon technological node while the distance between DQDs can be substantially larger (of the order of 100 nm) that much facilitates metallization and mitigates cross-talks.

## FUNDING

The work was carried out within the state assignment of NRC “Kurchatov institute.”

## CONFLICT OF INTEREST

The authors of this work declare that they have no conflicts of interest.